\documentclass[useAMS,usenatbib]{mn2e}
\usepackage{graphicx}
\usepackage{subfigure}

\newcommand{\Msun}{\mbox{$M_{\sun}$}}
\newcommand{\Mwd}{\mbox{$M_{\rm WD}$}}
\newcommand{\Rwd}{\mbox{$R_{\rm WD}$}}

\newcommand{\degree}{^{\scriptscriptstyle{\circ}}}
\newcommand{\gcms}{\mbox{$\rm g \, cm^{-2} \, s^{-1}$}}

\title[Compton Polarized X-rays in mCVs]{X-ray Polarization Signatures of Compton Scattering in Magnetic Cataclysmic Variables}
\author[McNamara et al.]
{A. L. McNamara$^{1}$\thanks{E-mail: aimee@physics.usyd.edu.au}, 
Z. Kuncic$^{1}$  
and K. Wu$^{2}$\\
$^{1}$School of Physics, University of Sydney, NSW 2006, Australia\\
$^{2}$Mullard Space Science Laboratory, University College London, Holmbury St
Mary, Surrey, RH5 6NT, UK}

\begin{document}

\date{Accepted . Received ; in original form }

\pagerange{\pageref{firstpage}--\pageref{lastpage}} \pubyear{2007}

\maketitle

\label{firstpage}

\begin{abstract}
Compton scattering within the accretion column of magnetic cataclysmic variables (mCVs) can induce a net polarization in the X-ray emission.   
We investigate this process using Monte Carlo simulations and find that significant polarization can arise as a result of the stratified flow structure in the shock-ionized column. We find that the degree of linear polarization can reach levels up to $\sim 8$\% for systems with high accretion rates and low white-dwarf masses, when viewed at large inclination angles with respect to the accretion column axis. These levels are substantially higher than previously predicted estimates using an accretion column model with uniform density and temperature. We also find that for systems with a relatively low-mass white dwarf accreting at a high accretion rate, the polarization properties may be insensitive to the magnetic field, since most of the scattering occurs at the base of the accretion column where the density structure is determined mainly by bremsstrahlung cooling instead of cyclotron cooling. 
\end{abstract}


\section{Introduction}
It is anticipated that X-ray polarimetry will provide us with a powerful method of probing the physical conditions and geometry of high-energy astrophysical systems. Accreting X-ray sources are expected to show a significant degree of polarization as a result of photon scattering in non-uniform distributions of matter in non-spherical geometries, such as accretion disks and columns \citep[see e.g.][]{Meszaros88, Rees75}.  

Accreting white dwarfs are strong X-ray sources during active states \citep[see e.g.][for reviews]{Kuulkers06, Wu03, Warner95}. In magnetized systems (the magnetic cataclysmic variables, mCVs), the accretion flow is confined by the magnetic field near the white dwarf. The supersonic accreting material becomes subsonic close to the white-dwarf surface, resulting in a standing shock, which ionizes and heats the plasma to temperatures  $kT\approx (10-40)$ keV, (where $k$ is Boltzmann's constant). The heated material in the post-shock flow cools by emitting bremsstrahlung X-rays and optical/IR cyclotron emission \citep{Lamb79, King79}. Bremsstrahlung radiation emitted by isotropic thermal electrons is not polarized. However, in strongly magnetic systems where cyclotron cooling is very efficient, Coulomb collisions might not be efficient enough to ensure an isotropic Maxwellian distribution for the electrons. Bremsstrahlung X-rays from such systems would be intrinsically polarized \citep[see e.g.][]{McMaster61}. For mCVs with a high accretion rate, the accretion column can have Thompson optical depths of up to a few, giving rise to substantial Comptonization signatures \citep*[see e.g.][]{Wu99,Kuncic05,McNamara07}. 

Our previous studies \citep{Kuncic05, McNamara07}, which used a nonlinear Monte Carlo algorithm for the simulations \citep{Cullen01a, Cullen01b},   demonstrates the substantial effects of Compton scattering on Fe K$\alpha$ emission lines in the post-shock flows of mCVs. Like the photons in the Fe lines, whose profiles are broadened and distorted by Compton scattering, the photons in the whole X-ray continuum can undergo multiple scatterings.  
Although Compton scattering would not readily introduce prominent spectral signatures in the $0.1 - 10$~keV continuum energy band, it can produce a net polarization due to the non-isotropic distribution of electrons in the accretion column, the lack of symmetry in the viewing geometry and perhaps the presence of the magnetic field in the accretion flow. A study by \cite{Matt04} has shown that the degree of polarization is $\simeq 4\%$ for a  cold, homogeneous and static accretion column. The degree of X-ray polarization in mCVs should be higher, given that the accretion flow is stratified and has non-zero temperature and velocity. It is expected that this polarization will be observable.  
   
In this paper, we investigate the X-ray polarization properties of mCVs by means of Monte Carlo simulations. We consider a more realistic model for the accretion column, which takes into account the full ionization structure of the post-shock column. The velocity, temperature and density profiles of the post-shock flow are derived using a model as described in \citet*{Wu94}. In Section 2, we outline the theory of polarized Compton scattering and the computational algorithm for the polarization calculation. In Sec. 3, we present our findings. A summary and conclusion are given in Sec. 4.

\section{Polarized Compton Scattering}
\subsection{Basic Physics}
The differential Klein-Nishina cross-section for the scattering of polarized electromagnetic waves off electrons is \citep[see][for example]{Heitler36, Jauch80}, 
\begin{equation}
  \frac{d\sigma}{d\Omega} = \frac{r_{0}^2}{2\gamma^2}\frac{X}{\left(1-\mu\beta\right)^2}\left(\frac{\nu'}{\nu}\right)^2 \qquad .
\end{equation}
Here, $r_{0}$ is the classical electron radius, $\gamma = (1-\beta^2)^{-1/2}$ is the Lorentz factor of electrons with velocity $v = \beta c$, $\mu$ is the cosine of the angle between the propagations of the incident photon and electron, and
\begin{eqnarray}
X &=& \frac{1}{2} \left(\frac{\kappa}{\kappa'} + \frac{\kappa'}{\kappa}\right) - 1   \label{fullX} \\
  && + 2 \left( \vec{e} \cdot \vec{e}\,' + \frac{\vec{e}\cdot\vec{p} \, \vec{e}\,'\cdot\vec{p}\,'}{\kappa} -  \frac{\vec{e}\,'\cdot\vec{p} \, \vec{e}\cdot\vec{p}\,'}{\kappa'} \right)  \nonumber
\end{eqnarray}  
where $\vec{p}$ is the electron momentum, $\vec{e}$ is the polarization vector of the incident photon and $\kappa = -\vec{p}\cdot\vec{k}$ and $\kappa'  = -\vec{p} \cdot \vec{k}\,'$, where $\vec{k}$ is the incident photon momentum. All other primed quantities denote values after scattering. Also,
\begin{equation}
  \frac{\nu'}{\nu} = (1 - \mu\beta)\left[(1 - \mu'\beta) + \frac{h\nu}{\gamma m_{\rm e} c^2}(1-\cos\theta)\right]^{-1} 
\end{equation}
where $\nu$ is the initial frequency of the photon and $\theta$ is the scattering angle (angle between the incident and scattered photon propagations), and $\mu'$ is the cosine of the angle between the propagations of the scattered photon and the incident electron. 

The expression for $X$ given by (\ref{fullX}) simplifies considerably by specializing to the rest frame of the incident electron. In this case, 
$p = (m_{\rm e}c^2,0,0,0)$ and a specific gauge can be chosen such that $\vec{e}_{0} =\vec{e}_{0}\,' = 0$, then $\vec{p} \cdot \vec{e} = \vec{p} \cdot \vec{e}\,'= 0 $ and (\ref{fullX}) becomes \citep[see e.g.][for details]{Jauch80}
\begin{equation}
  X_{\rm e} = \frac{1}{2}\left(\eta_{\rm e} + \eta_{\rm e}^{-1}\right)_{\rm e} - 1 + 2\left({\vec{e}} \cdot \vec{e}\,'\right)^2 
\end{equation}
where the subscript `e' denotes a quantity calculated in the electron rest frame and 
\begin{equation}
 \eta_{\rm e} \equiv \left(\frac{\nu'}{\nu}\right)_{\rm e}  = \left[1 + \frac{h\nu_{\rm e}}{m_{\rm e} c^2}(1-\cos\theta_{\rm e}) \right]^{-1} \qquad .
\end{equation}
The differential cross section in the electron rest frame simplifies to, 
\begin{equation}
  \left(\frac{d\sigma}{d\Omega}\right)_{\rm e} = \frac{1}{4} r_{0}^2 X_{\rm e} \eta_{\rm e}^2  \qquad . 
\end{equation}

We can consider the scattered radiation as composed of linearly polarized components perpendicular and parallel to the incident plane of polarization. Then the differential cross section for polarized scattering in the electron rest frame can be written in the more familiar form 
\begin{equation}
  \left(\frac{d\sigma}{d\Omega}\right)_{\rm e} = \frac{1}{2}r_{0}^2\eta^2_{\rm e}  \left[ \eta_{\rm e} + \eta_{\rm e}^{-1} -2 \sin^2 \theta_{\rm e} \cos^2 \phi_{\rm e} \right] \qquad .
  \label{P}
 \end{equation}

The polarization vector $\vec{e}\,'$ for the polarized fraction $P$ of photons is perpendicular to the scattering plane and is defined by \citep{Angel69},
\begin{equation}
  \vec{e}\,' = \frac{1}{|\vec{e}\,'|} (\vec{e} \times \hat{\Omega}') \times \hat{\Omega}' 
  \label{Angeleqn}
\end{equation}
where $\hat{\Omega}'$ is the propagation directional unit vector of the scattered photon. For the remaining $1-P$ fraction of photons $\vec{e}\,'$ is randomly distributed in the plane perpendicular to $\hat{\Omega}'$. In the case where the radiation is intially unpolarized, the degree of linear polarization induced by Compton scattering is given by \citep[see e.g.][]{Dolan67}, 
\begin{equation} 
  P = \sin^2\theta_{\rm e} \left( \eta_{\rm e} + \eta^{-1}_{\rm e} -1 + \cos^2\theta_{\rm e}\right)^{-1}   \quad .
\end{equation}

\begin{figure}
	\begin{center}
	\includegraphics[width=9.0truecm]{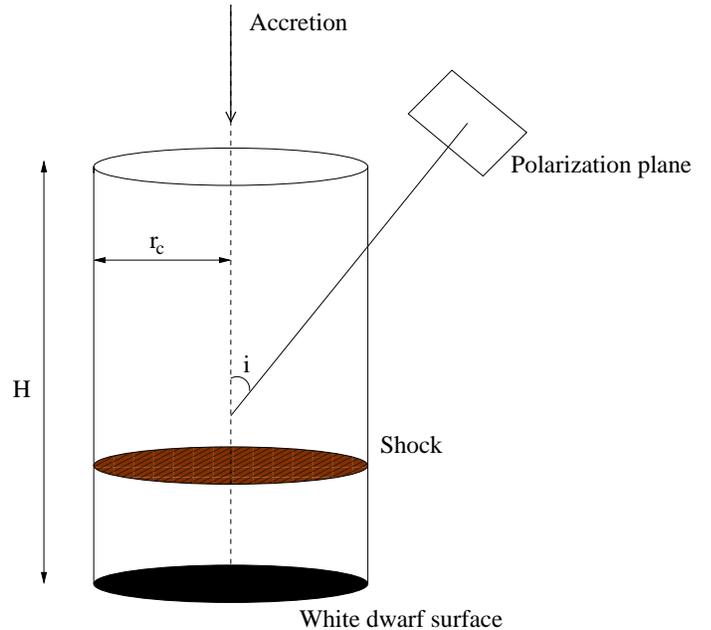}
	\caption{Schematic geometry of an mCV accretion column with total height H and cross-sectional radius $r_{\rm c}$, viewed from an inclination angle $i$.} 
	\label{column}
	\end{center}
\end{figure}

\subsection{Computational Algorithm}

\citet*{Poz83} outlined a Monte Carlo algorithm to model Compton scattering of photons. The algorithm was implemented into a non-linear code developed by \citet{Cullen01a} which was subsequently used to investigate Compton scattering of Fe lines in mCVs \citep{Kuncic05,McNamara07}. 
Here, we generalize the algorithm to include the polarization. If polarization is not included, the azimuthal angle $\phi$ of the scattered photons may be sampled uniformly. However, in polarization calculations the azimuthal distribution of the scattered photons depends on the polarization vector of the incident photons and is anisotropic, thus uniform sampling is not applicable. In the algorithm for our previous calculations, the energy and direction of the scattered photons are calculated in the white-dwarf frame. Polarized scattering in the white-dwarf frame, however, requires sampling multivariate distributions of the scattered photon energy $h\nu'$, the cosine of the angle between the scattered photon and electron $\mu'$, the azimuthal angle $\phi$ and the angles between the polarization vectors and photon propagation vectors. To avoid this complication, a Lorentz transformation into the incident electron rest frame is performed. The Lorentz transformation of the photon momentum and energy between the electron rest frame and the lab frame is given by 
\begin{equation}
  \vec{k}_{\rm e}=\vec{k}-k\left[\gamma\nu-(\gamma-1)\hat{k}\cdot\hat{\beta}\right]\hat{\beta} 
  \label{momlorentz}
\end{equation}  
and
\begin{equation}
  \nu_{\rm e} = \nu\gamma(1-\beta \cos\theta)
\end{equation}  
respectively. Once the resulting energy and momentum are calculated in the electron rest frame, a transformation is made into the lab frame using (\ref{momlorentz}) with a reversed $\hat{\beta}$. The resulting scattering angle is transformed using
\begin{equation}
  \cos \theta = \frac{\cos\theta_{\rm e} + \beta}{1+\beta\cos\theta_{\rm e} } \qquad .
\end{equation}   

In the incident electron rest frame we sample the variables $\mu'_{\rm e}$ and $\phi_{\rm e}$ using the inverse-function method and devise a rejection technique. We follow the same procedure outlined in \citet{Sobol79} and \citet{Poz83} to determine the joint distribution density of the  variables,
\begin{equation}
  \Pi(\mu'_{\rm e},\phi_{\rm e}) = \sigma^{-1}\left(\frac{d\sigma}{d\Omega}\right)_{\rm e} \sim X_{\rm e} \left(\frac{\nu'}{\nu}\right)_{\rm e}^2 \qquad .
\end{equation}
We rewrite the joint density in the form \citep{Sobol79}
\begin{equation}
  \Pi(\mu'_{\rm e},\phi_{\rm e}) = \sigma^{-1}\Pi_1(\mu'_{\rm e},\phi_{\rm e})Y
\end{equation}
where the normalized density $\Pi_1(\mu'_{\rm e},\phi_{\rm e}) =\frac{1}{2\pi} \cos^2\phi_{\rm e}$ and
\begin{equation}
  Y = \left(\frac{\nu'}{\nu}\right)^2_{\rm e} X_{\rm e} \leq 2   \qquad .
  \label{Y}
\end{equation}

Our algorithm is as follows:  
\begin{enumerate}
  \item In the white-dwarf frame, choose the incident electron energy $E$ and momentum $\vec{p}$ from a Maxwellian distribution.
  \item Calculate $\mu$, $\beta$ and $\gamma$.
  \item Calculate the total cross-section $\sigma$. 
  \item Determine whether the scattering event will occur using a rejection algorithm \citep{Cullen01a,Cullen01b}. If the scattering event is  rejected, then the photon continues in the same direction without scattering.
  \item If scattering is accepted, then
    \begin{enumerate}
      	\item Lorentz transform the incident photon momentum $\vec{k}$, energy $h\nu$ and $\mu$ into the electron rest frame. 
      	\item Select two random numbers $u_1$ and $u_2$, and calculate a possible 
	      direction of scattering for the density $\Pi_1(\mu'_{\rm e},\phi_{\rm e})$ by solving:  
	      \begin{eqnarray}
	         &&\mu_{\rm e}' = 2u_1 - 1  \\   \nonumber
	         &&2\phi_{\rm e} + \sin(2\phi_{\rm e}) = 4\pi u_2 
	      \end{eqnarray}
       \item Calculate the scattered photon direction $\hat{\Omega}'_{\rm e}$ and scattering angle $\theta_{\rm e}$ from $\hat{\Omega}_{\rm e} \cdot \hat{\Omega}_{\rm e}'$.
       \item Determine the energy of the scattered photon $h\nu'_{\rm e}$ from,
      	     \begin{equation}
	        \frac{\nu_{\rm e}'}{\nu_{\rm e}} = \left[1+(h\nu_{\rm e}/m_{\rm e}c^2)(1-\cos_{\rm e}\theta_{\rm e})\right]^{-1}
	     \end{equation}
       \item Determine $Y$ from (\ref{Y}). If $Y \leq 2$, accept the scattered quantities, else 
	     recalculate $\mu'_{\rm e}$ and $\phi_{\rm e}$.
       \item Once $\mu'_{\rm e}$ and $\phi_{\rm e}$ are known, determine whether the scattered photon will be polarized by calculating the degree of polarization $P$ and extracting a random number $u_3$. If $u_3 < P$, then 
	     the photon is unpolarized else the new polarization vector $\vec{e}\,'_{\rm e}$ is chosen from~(\ref{Angeleqn}).
       \item Lorentz transform the scattered photon momentum $\vec{k'}_{\rm e}$, energy $h\nu'_{\rm e}$ and scattering angle $\theta_{\rm e}$ into the 
       white-dwarf frame.
       \item Calculate the total linear polarization $P$ of photons emerging in a specified direction as follows: define a viewing plane and calculate the Stokes parameters $Q$ and $U$ by projecting $\vec{e}$ onto the plane \citep{Matt96}. Then calculate $P = \sqrt{Q^2 + U^2}/I$, where $I$ is the number of photons scattered in the chosen direction.
     \end{enumerate}	
\end{enumerate} 

In this paper, we only consider unpolarized incident photons and the initial polarization vector $\vec{e}$ is chosen randomly in the plane perpendicular to the photon propagation direction $\hat{\Omega}$, i.e. with the condition $\vec{e} \cdot \hat{\Omega} = 0$. The energy of the incident photons is sampled from a bremsstrahlung spectral function $N(\nu)d\nu$ using rejection methods as described in \citet{Press92}. Since the distribution function $N(\nu)d\nu$ is too complex to sample directly, we consider an approximate spectral function $F(\nu)d\nu$ which lies above $N(\nu)d\nu$ everywhere for a predefined energy range $h\nu_{\rm max} \leq h\nu \leq h\nu_{\rm min}$. We use the comparison function from \citet{Fromerth01} for the thermal bremsstrahlung photon distribution function with cutoff temperature $T$,
\begin{equation}
  f_{\rm tb} = \left\{ \begin{array}{ll}
  x^{-1.2} &\mbox{for $h\nu_{\rm min}/kT \leq x < 1$,} \\ 
  e^{-x}   &\mbox{for $x > 1$.}
       \end{array} \right.
 \end{equation}
 where $x=h\nu/kT$. The comparison function is sampled to determine the energy of a photon $h\nu$. If $h\nu$ lies in the predefined energy range and $uf(\nu) \leq N(\nu)$, where $u$ is a standard deviate, then the sampled energy is accepted. If the conditions are not satisfied then we sample a new energy from the comparison function. 
 
\section{Results and Discussion}

\begin{figure}
	\includegraphics[width=8.5truecm]{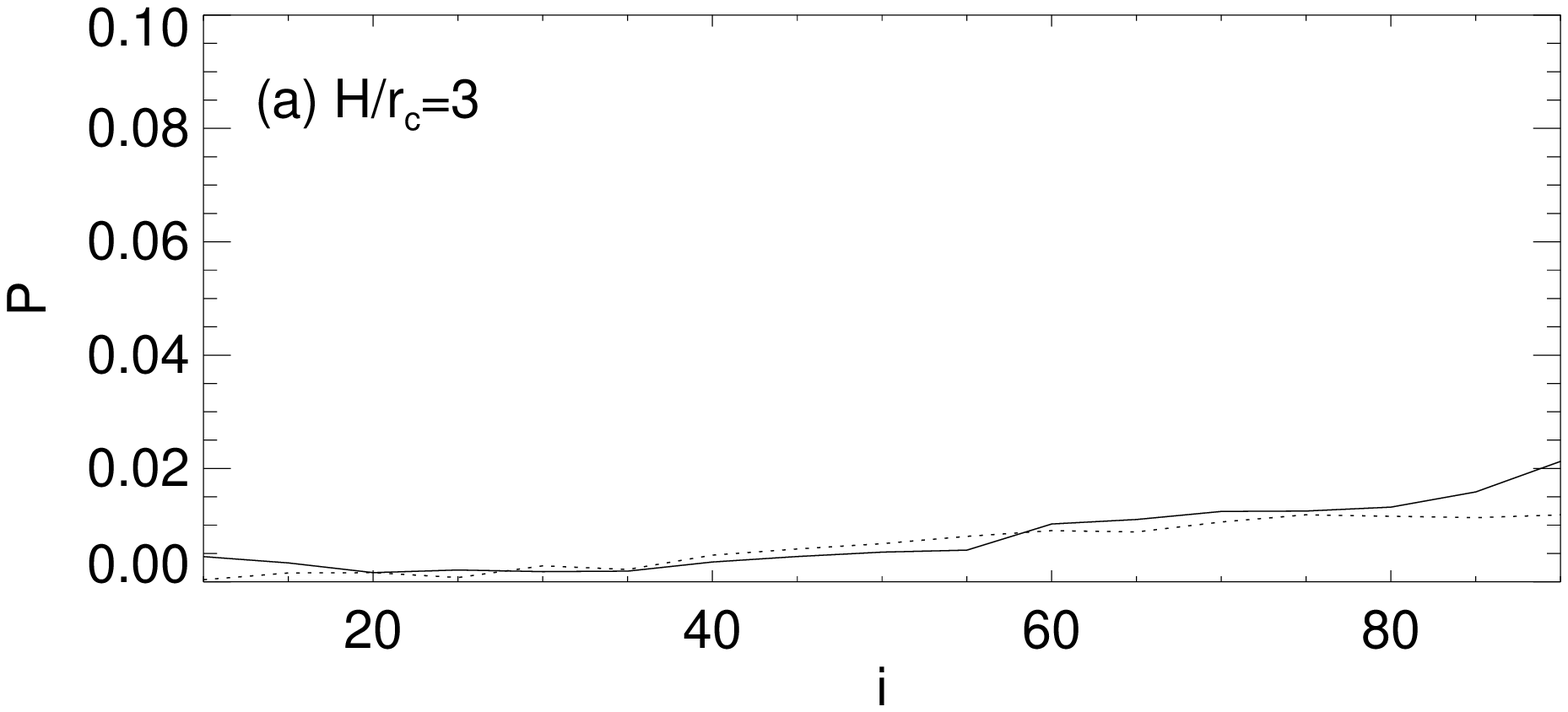} \\
	\includegraphics[width=8.5truecm]{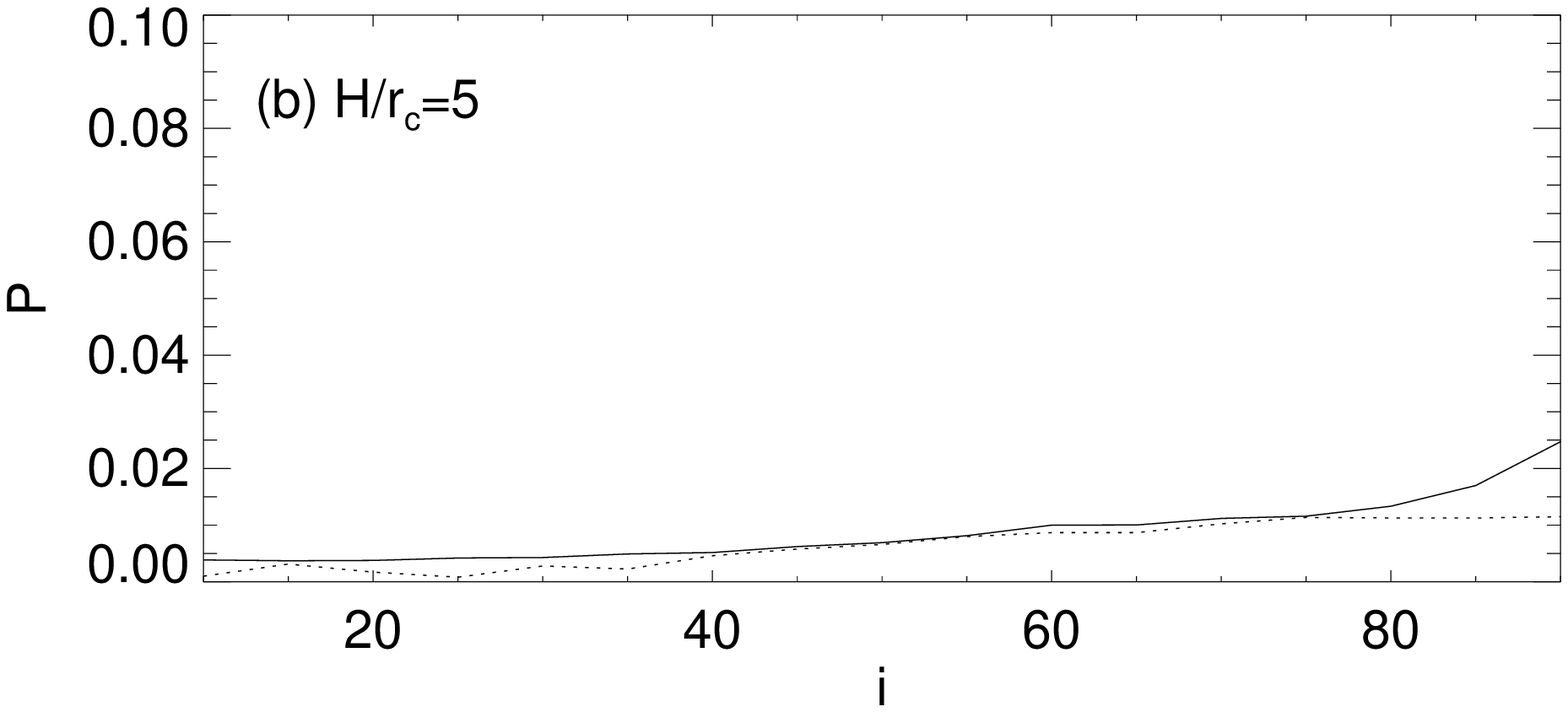} \\
	\includegraphics[width=8.5truecm]{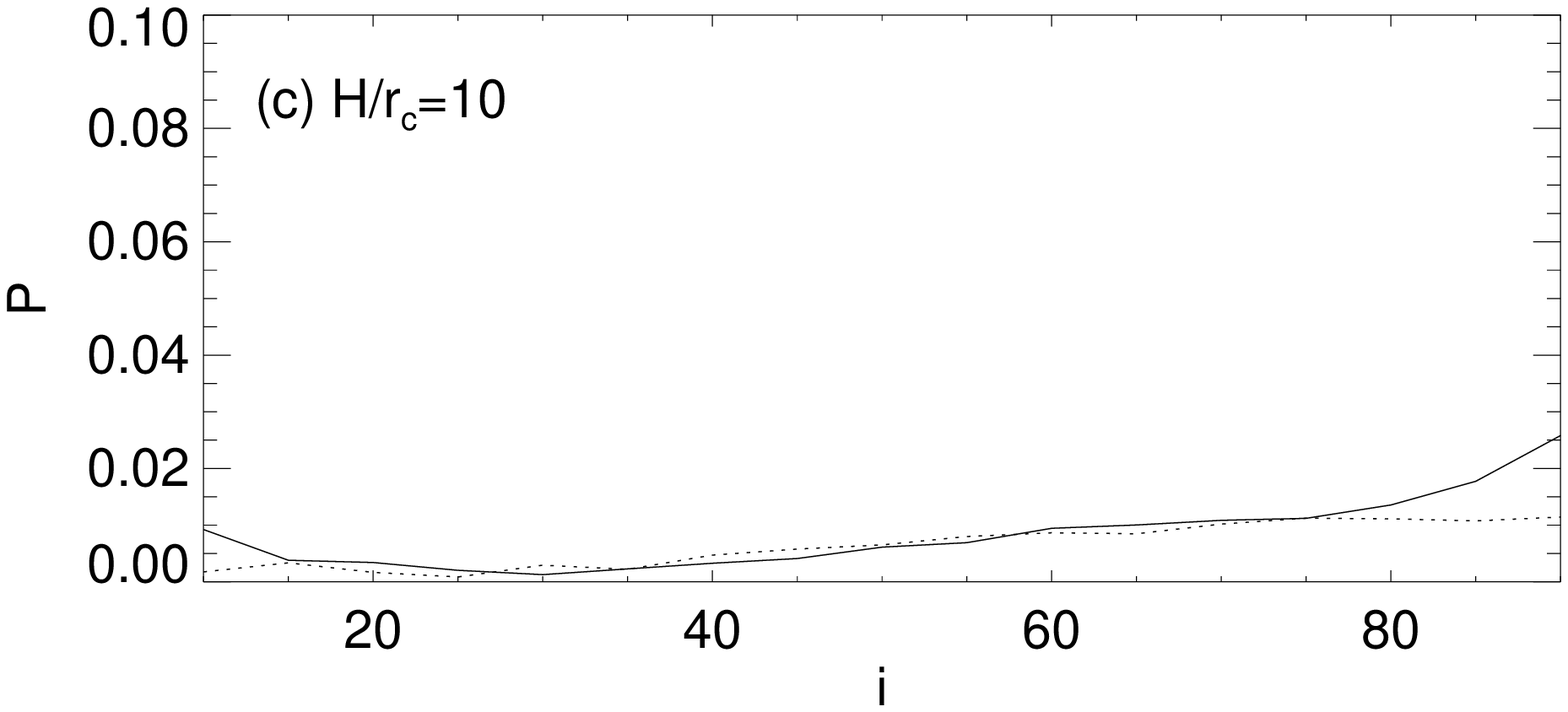} 
	\caption{Polarization degree $P$ plotted as a function of the inclination angle $i$ for an accretion column with a white-dwarf mass $\Mwd = 1.0 \Msun$,  specific accretion rates of $\dot m = 1.0 \, \gcms$ (dotted line) and $\dot m = 10 \, \gcms$ (solid line) and for different column height-to-radius ratios $H/r_{\rm c}$, as shown. The optical depth across the column at the shock is $\tau = 0.3$ and $0.04$ for the high and low $\dot m$ cases, respectively.}
\label{M1}
\end{figure}

\begin{figure}
	\includegraphics[width=8.5truecm]{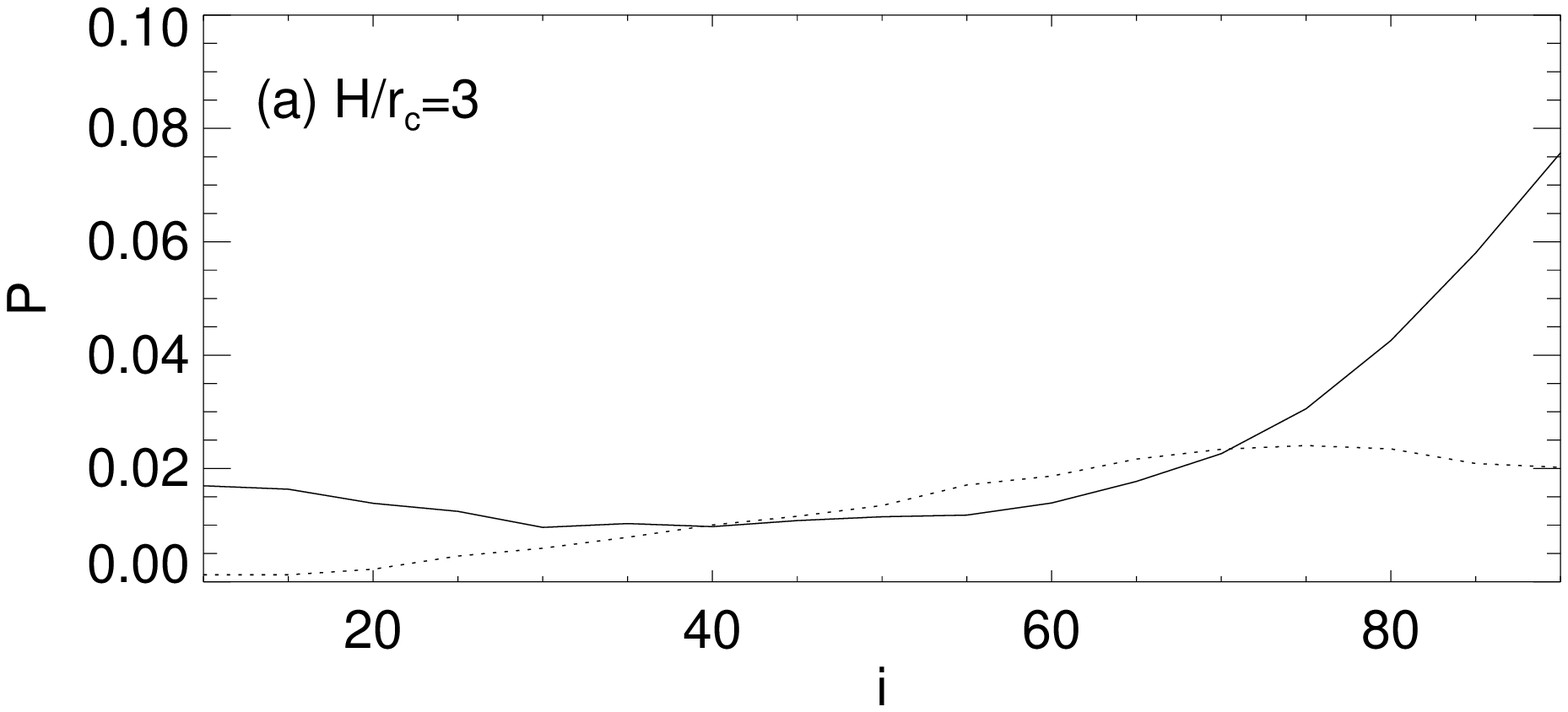} \\
	\includegraphics[width=8.5truecm]{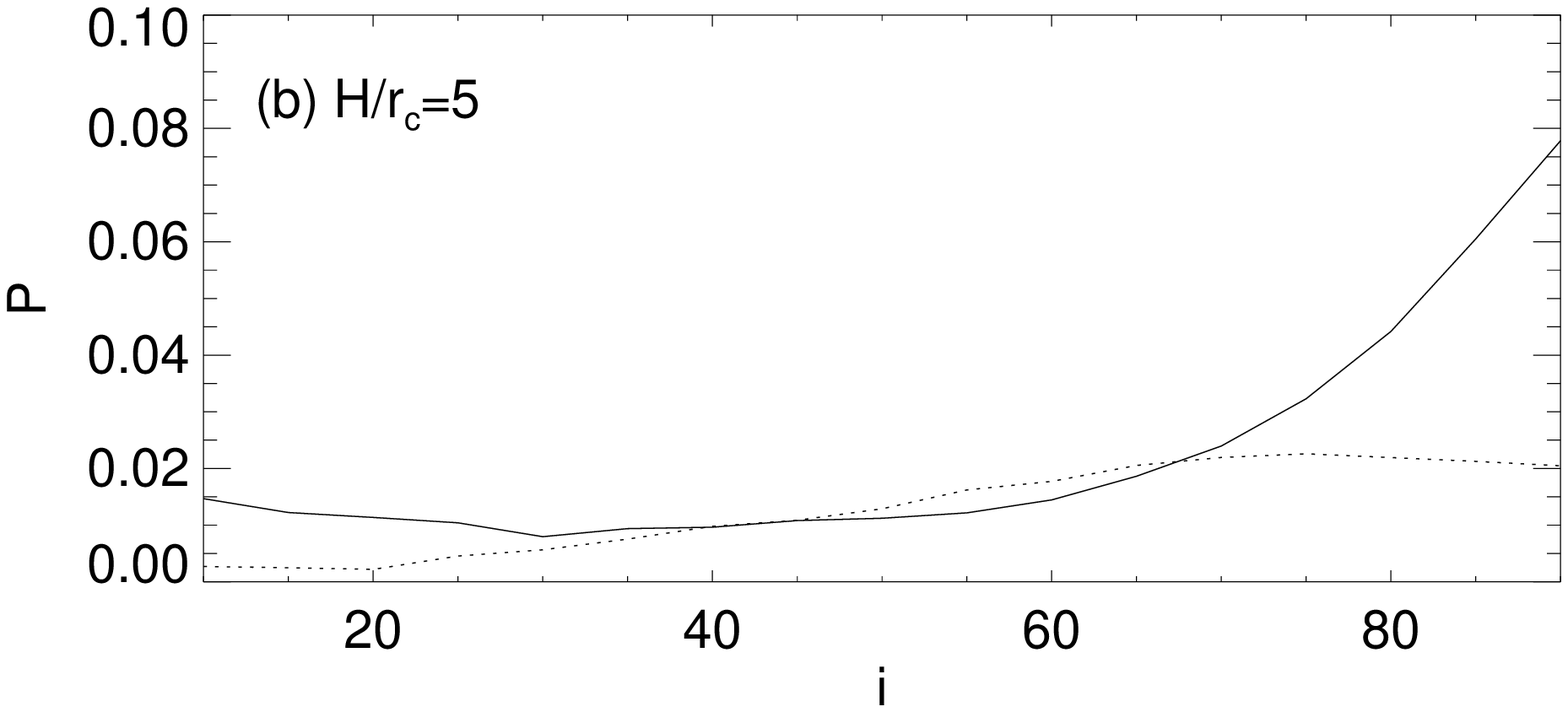} \\
	\includegraphics[width=8.5truecm]{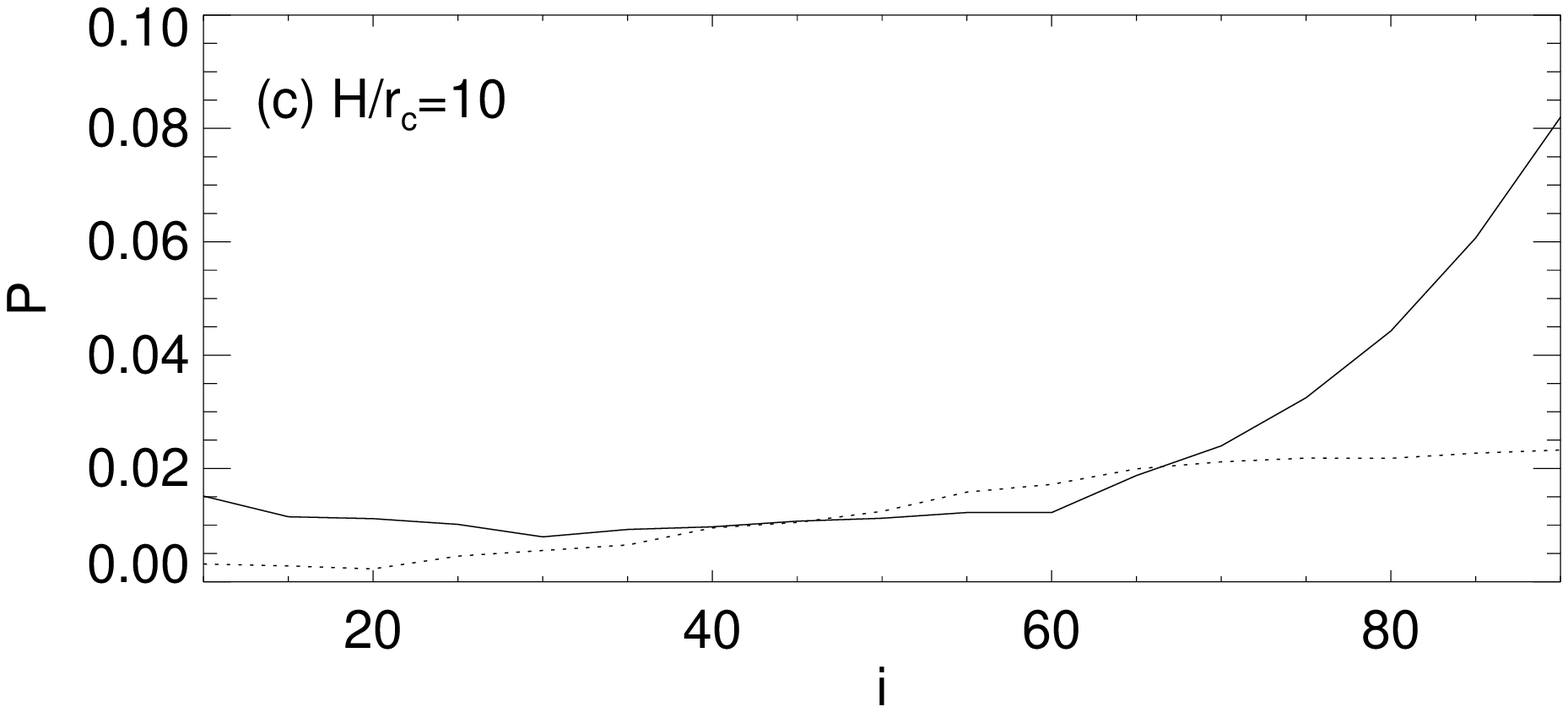}
	\caption{Polarization degree $P$ plotted as a function of the inclination angle $i$ for an accretion column with a white-dwarf mass $\Mwd = 0.5 \,\Msun$,  specific accretion rates of $\dot m = 1.0 \, \gcms$ (dotted line) and $\dot m = 10 \, \gcms$ (solid line) and for different column height-to-radius ratios $H/r_{\rm c}$, as shown. The optical depth across the column at the shock is $\tau = 1.0$ and $0.1$ for the high and low $\dot m$ cases, respectively.}
\label{M05}
\end{figure}

\begin{figure*}
	\includegraphics[width=15.0truecm]{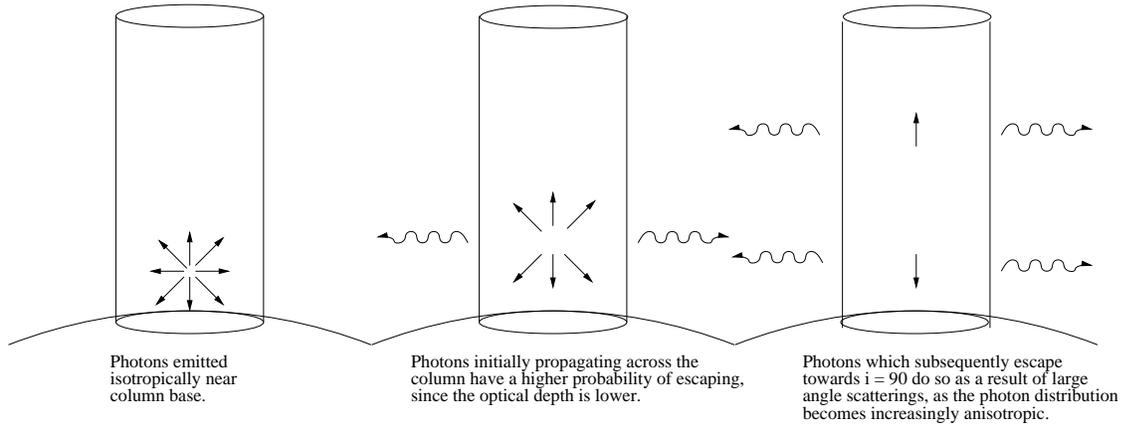}
	\caption{Propagation of an isotropic distribution of photons injected at the base of the mCV accretion column. As the photons propagate through the length of the column their distribution becomes increasingly anisotropic.}
\label{photonprop}
\end{figure*}	

\begin{figure}
	\includegraphics[width=8.5truecm]{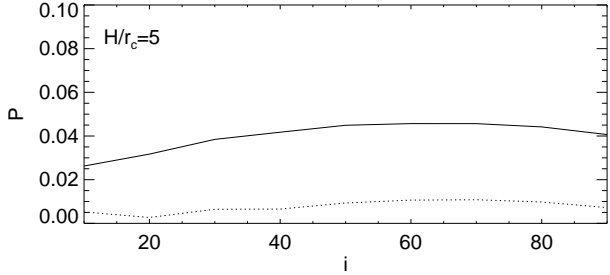} 
	\caption{Same as Fig.~\ref{M05} (b), except for a uniform density accretion column}
\label{uniform}
\end{figure}

\begin{figure}
	\includegraphics[width=8.5truecm]{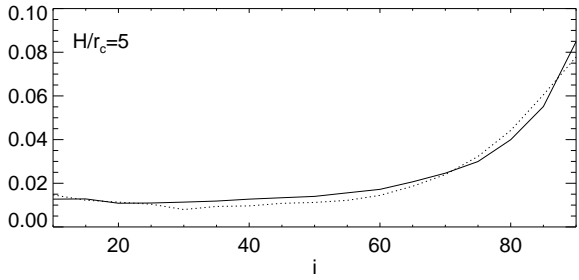}
	\caption{Polarization degree $P$ plotted as a function of inclination angle $i$ for an mCV with a white-dwarf mass $\Mwd = 0.5 \,\Msun$ and specific accretion rate $\dot m = 10 \, \gcms$ with a ratio of cyclotron to bremsstrahlung cooling $\epsilon_{\rm s} = 10$ (solid curve) and $\epsilon_{\rm s} = 0$ (dotted curve).}
\label{cyclotron}
\end{figure}     

\begin{figure}
	\includegraphics[width=8.5truecm]{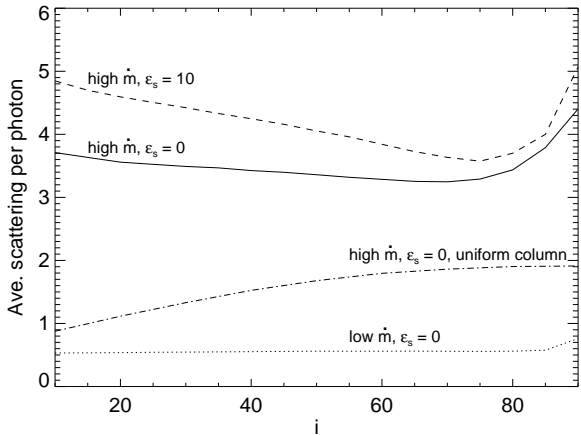}
	\caption{The average number of scatterings per photon plotted as a function of inclination angle $i$ for a white-dwarf mass $\Mwd = 0.5 \, \Msun$ with $H/r_{\rm c} = 5$.  The solid and dotted line shows the average scattering per photon for $\dot m = 10$ and $1 \, \gcms$, respectively (c.f. Fig.~\ref{M05}b). The dashed line shows the results for $\dot m = 10 \, \gcms$ when $\epsilon_{\rm s} = 10$ (c.f. Fig.~\ref{cyclotron}). The dot-dash line shows the average number of scattering per photon for a uniform column with $\dot m = 10 \, \gcms$ (c.f. Fig.~\ref{uniform}). }
\label{scatnum}
\end{figure}

\begin{figure}
	\includegraphics[width=8.5truecm]{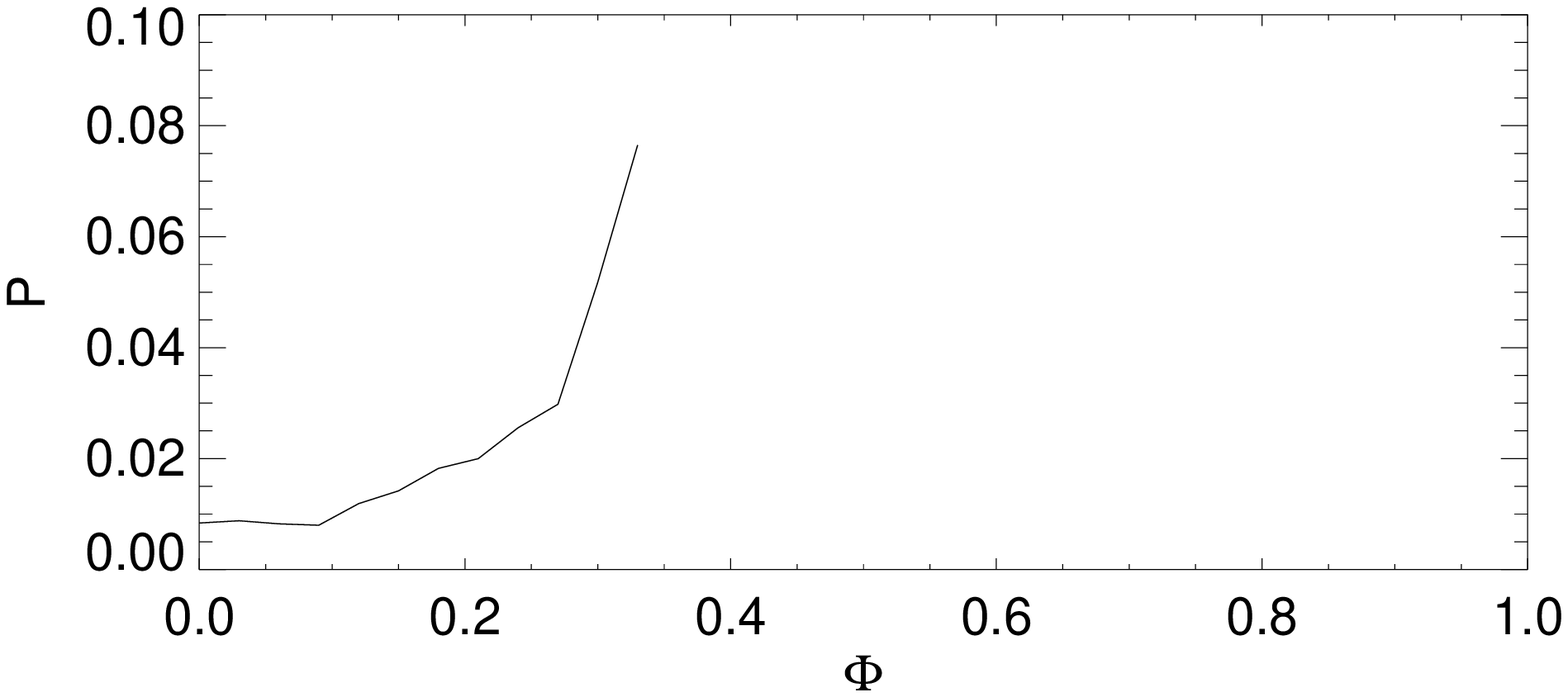}
	\caption{Linear polarization $P$ plotted as a function of phase angle $\Phi$ for the mCV GK Per with a white-dwarf mass $\Mwd = 0.63 \,\Msun$ and specific accretion rate $\dot m = 10 \, \gcms$. }
\label{GKPer}
\end{figure}

Our model scattering region is a cylindrical column, which is divided into a heated post-shock flow and a cool pre-shock flow. Since the upstream infall is supersonic and close to free fall, we assume that the velocity and density are constant in the pre-shock region. The post-shock flow is stratified in density, velocity and temperature, and these profiles are calculated using the hydrodynamic model described in \citet{Wu94}.  
The relative efficiency of cyclotron and bremsstrahlung cooling is determined by the magnetic field, through an efficiency parameter $\epsilon_{\rm s}$ evaluated at the shock. The ionization structure of the post-shock region is dependent on the white-dwarf mass $\Mwd$, and radius $\Rwd$, the specific mass accretion rate $\dot m$ and the ratio of the efficiencies of cyclotron to bremsstrahlung cooling $\epsilon_{\rm s}$ \citep[see][for futher details]{Kuncic05, McNamara07}. In this study, we consider cases where $\epsilon_{\rm s} = 0$ and $10$. We also assume that the emitted bremsstrahlung photons are intially unpolarized and are emitted from the base of the column where the emissivity peaks \citep{Wu94a}.

We consider an accretion column for two different white-dwarf mass-radius values: $\Mwd = 0.5 \, \Msun$, $\Rwd = 9.2 \times 10^8$ cm and $\Mwd = 1.0 \, \Msun$, $\Rwd = 5.5 \times 10^8$ cm \citep{Nauenberg72}. For each mass, we consider two different specific mass accretion rates: $\dot m = 1 \,\gcms$ and $\dot m = 10 \,\gcms$, roughly corresponding to the low and high ends of typical accretion rates of mCVs. The viewing inclination angle $i$ is measured from the cylinder axis (see Figure~\ref{column}). A polarization plane is defined for a small range of inclination angle $i$ such that $\Delta i \approx 5\degree$. In the simulations, each photon is followed until it escapes the column. If a photon strikes the polarization plane, it is binned and contributes to the overall polarization of the emerging beam for that particular $i$.   

The degree of polarization is very sensitive to the geometry and density structure in the accretion column. For a lower $\Mwd$, corresponding to a larger $\Rwd$, the optical depth across the column is higher for a fixed $\dot m$ and thus, photons undergo more scatterings on average. In the case of multiple scatterings, the resulting polarization is largely determined by the conditions at the last scattering surface. We consider a range of different column geometries by varying the column height, $H$. The column radius $r_{\rm c}$ is fixed to $0.1\,\Rwd$.   

Figures~\ref{M1} and \ref{M05} show the linear polarization $P$ plotted as a function of $i$ for white-dwarf masses $\Mwd = 1.0 \, \Msun$ and $\Mwd =0.5 \, \Msun $, respectively with $\epsilon_{\rm s} = 0$. The solid curves are for $\dot m = 10 \,\gcms$ and the dotted curves are for $\dot m = 1 \,\gcms$. We consider three different column geometries: $H/r_{\rm c}$ = 3, 5 and 10. In each case, $10^{9}$ photons were injected and emitted isotropically at the base of the accretion column where the density and hence, emissivity peaks. We define $P$ as positive when the polarization is perpendicular to the projection of the cylinder axis onto the projection plane, as is the convention for axisymmetric geometries. 

In all cases, $P$ increases with $\dot m$. This is because the electron number density and hence, scattering probability increases with $\dot m$. Indeed, in all the low $\dot m$ cases, the average number of scatterings is $\leq 1$, whereas photons scatter on average more than once in all the high $\dot m$ cases (see Fig.~\ref{scatnum}). The scattering probability is also slightly higher for larger $H/r_{\rm c}$, producing higher $P$ for these column geometries. Figs.~\ref{M1} and \ref{M05} also reveal that $P$ increases with $i$. This trend arises because photons in the initial seed distribution at the column base escape quickly with none or very few scatterings across the radius of the column ($i\approx90\degree$ direction), leaving behind a photon distribution that becomes increasingly more anisotropic (see Fig.~\ref{photonprop}). Photons that now escape in directions $i \approx 90\degree$ must do so through large-angle scatterings which result in a high net $P$. Conversely, photons do not need to undergo large-angle scatterings in order to escape in directions $i \approx 0$, hence $P$ is lowest for these viewing angles.

The steep rise in $P$ towards $i\approx90\degree$, reaching values up to 8\% in the high $\dot m$ cases, differs considerably from the results of \citet{Matt04}, which show a gradual rise in $P$ with $i$ towards a maximum of 4\% at $i\approx 90\degree$. This difference can be attributed to the assumption of a uniform column in Matt's model. To demostrate this difference, in Fig.~\ref{uniform} we plot $P$ vs. $i$ for a uniform column with the same parameters as those used for Fig.~\ref{M05}b. In this case, the photons are emitted uniformly throughout the column. The resulting $P$ at $i\approx 90\degree$ is substantially lower than that predicted for the nonuniform column (where the photons are emitted at the base of the column). This is because the photon distribution remains quasi-isotropic throughout the uniform column and thus, fewer photons undergo large angle scatterings to escape in the direction $i\approx 90\degree$. Fig.~\ref{scatnum} shows the average number of scatterings each photon undergoes before escaping the column at a particular $i$. In the case of a uniform column the average number of scatterings is lower than the non-uniform case for the the same $\dot m$. 

We also investigate the effect of cyclotron cooling dominated accretion flows on polarization. Fig.~\ref{cyclotron} compares the polarization degrees for an mCV with $\Mwd = 0.5 \, \Msun$, $\dot m = 10 \, \gcms$ and $H/r_{\rm c} = 5$ when cyclotron cooling at the shock is negligible ($\epsilon_{\rm s} = 0$,  dotted curve) and when it dominates ($\epsilon_{\rm s} = 10$, solid curve). Interestingly, there is not much difference between these two cases. This can be understood as follows. The presence of cyclotron cooling makes the accretion column more compact and the density enhancement increases the average number of photon scatterings (Fig.~\ref{scatnum}). However, for a sufficiently high accretion rate, the column will be optically thick regardless of whether bremsstrahlung cooling or cyclotron cooling dominates at the shock. For multiple scatterings, the resulting polarization degree is determined largely by the final scattering. Generally in cyclotron dominated flows, the last scattering surface will be in the dense bremsstrahlung cooling zone close to the base of the column, where the physical conditions are very similar to the bremsstrahlung cooling dominated case. In both cases, the scattering electrons are relatively cool with low velocities, and as a result aberration is negligible and Compton recoil will dominate. With three or four scatterings beforehand, incident photon directions for a subsequent final scattering will be approximately randomized and hence unpolarized. Thus, the average polarization $P$ of photons emerging from the last scattering surface depends only on the scattering angle (and hence the viewing inclination) and is insensitive to cyclotron cooling. 

The situation may be different for lower accretion rates.  If the post-shock region becomes semi-transparent to scattering, photons can emerge from the entire structure of the post-shock flow. In that case, cyclotron cooling, and by implication the white-dwarf magnetic field which modifies the flow, can affect the polarization. We may thus conclude that polarization is a robust diagnostic of geometry and orientation of systems if they have sufficiently high accretion rates.  

Finally, we show an application to an mCV with high accretion rate and low magnetic field. We construct a model accretion column using mCV parameters similar to that of the intermediate polar GK Per (Fig.~\ref{GKPer}): $\Mwd = 0.63 \, \Msun$ and $\dot m = 10 \, \gcms$ \citep{Morales02, Vrielmann05}. The predicted fractional polarization increases towards $\Phi = 0.33$ where $P$ peaks at $\approx 0.075$. Whether the predicted $P$ and its phase-dependent variation could be observed would depend upon, among other things, the extent of dilution by background emission. Note that the predicted $P$ for GK Per type mCVs would be higher as the accretion flow resembles a curtain rather than a cylinder. Also, X-rays reflected by dense material further upstream in the accretion disk and the disk magnetosphere coupling region will contribute to the observed polarization. Note that for very strong field systems with low accretion rates (e.g. polars in an intermediate accretion state), an anisotropic temperature distribution due to thermal decoupling between charged particles at the shock \citep[see][]{Saxton05} may complicate the situation by imprinting an intrinsic polarization on the seed bremsstrahlung radiation. \citep[see][for a review of polars and intermediate polars]{Warner95}. Also, an additional polarized Compton component resulting from reflection off the white dwarf surface can contribute to the overall spectrum above a few keV and could introduce an energy dependence on the polarization \citep{Matt04}.

\section{Summary and Conclusion}
We have investigated the properties of Compton polarized X-rays in the accretion column of mCVs using Monte Carlo simulations. The calculations take into account a post-shock region stratified in temperature, density and velocity. The degree of linear polarization for scattered X-rays was calculated for a range of different column geometries, white dwarf masses and accretion rates. 

We have found that the resulting polarization is sensitive to the density structure in addition to the viewing geometry. The non-uniform density structure in the post-shock column has a significant effect on the photon distribution and average number of scatterings. We have demonstrated that enhanced emissivity near the base of the column results in photon distributions that become increasingly more anisotropic throughout the column length so that X-rays escaping at large angles with respect to the column axis can do so only through increasingly larger angle scatterings. This produces more strongly polarized X-rays emerging from directions perpendicular to the column axis than directions parallel to it. 

We find that the degree of polarization is two times larger than the levels $\simeq$ 4\% predicted in previous studies, for a cold, static accretion column with uniform density. We demonstrate that for usual mCV parameters (e.g.\ $\Mwd = 0.5 \Msun$ and $\dot m = 10 \, \gcms$), the degree of polarization can reach up to $\simeq 8$\% for accretion columns with high viewing inclination angles. The polarization is not significantly affected by introducing cyclotron cooling, provided that the accretion rate of the system is sufficiently high.


\bsp

\label{lastpage}


\begin{thebibliography}{99}
\bibitem[\protect\citeauthoryear{Angel}{1969}]{Angel69} Angel J.R.P., 1969, ApJ, 158, 219
\bibitem[\protect\citeauthoryear{Cullen}{2001a}]{Cullen01a} Cullen J.G. 2001a, PhD Thesis, University of Sydney
\bibitem[\protect\citeauthoryear{Cullen}{2001b}]{Cullen01b} Cullen J.G. 2001b, JCoPh, 173, 175 
\bibitem[\protect\citeauthoryear{Cropper}{1990}]{Cropper90} Cropper M., 1990, Space Science Reviews., 54, 195
\bibitem[\protect\citeauthoryear{Dolan}{1967}]{Dolan67} Dolan J.F., 1967, Space Science Reviews, 6, 579  
\bibitem[\protect\citeauthoryear{Fromerth, Melia \& Leahy}{Fromerth et al.}{2001}]{Fromerth01} Fromerth M.J., Melia F., Leahy, D.A., 2001, ApJ, 547, L129 
\bibitem[\protect\citeauthoryear{Heitler}{1936}]{Heitler36} Heitler W., 1936, Quantum Theory of Radiation, Oxford University Press, p. 146
\bibitem[\protect\citeauthoryear{Hua}{1997}]{Hua97} Hua X.-M., 1997, Computers in Physics, 11, 6 
\bibitem[\protect\citeauthoryear{Jauch \& Rohrlich}{1980}]{Jauch80} Jauch J.M., Rohrlich F., 1980, The Theory of Photons and Electrons, Springer-Verlag, p. 229
\bibitem[\protect\citeauthoryear{King \& Lasota}{1979}]{King79} King A.R., Lasota J.P., 1979, MNRAS, 188, 653
\bibitem[\protect\citeauthoryear{Kuncic, Wu \& Cullen}{Kuncic et al.}{2005}]{Kuncic05} Kuncic Z., Wu K., Cullen, J., 2005, PASA, 22, 56
\bibitem[\protect\citeauthoryear{Kuulkers et al.}{2006}]{Kuulkers06} Kuulkers E., Schwope A., Norton A., Warner B., 2006, in Lewin W.H.G., van der Klis M., eds, Compact Stellar X-ray Sources, Cambridge Astrophysics Series 39, p421
\bibitem[\protect\citeauthoryear{Lamb \& Masters}{1979}]{Lamb79} Lamb D.Q., Masters A.R., 1979, ApJ, 234, L117
\bibitem[\protect\citeauthoryear{Matt}{2004}]{Matt04} Matt G., 2004, A\&A, 423, 495
\bibitem[\protect\citeauthoryear{Matt et al.}{1996}]{Matt96} Matt G., Feroci M., Rapisarda M., Costa E., 1996, Rad. Phys. Chem., 48/4, 403 
\bibitem[\protect\citeauthoryear{McMaster}{1961}]{McMaster61} McMaster W.H., 1961, Rev. Mod. Phys., 33, 8
\bibitem[\protect\citeauthoryear{McNamara et al.}{2008}]{McNamara07} McNamara A.L., Kuncic Z., Wu K., Galloway D.K, Cullen J., 2008, MNRAS, 383, 962
\bibitem[\protect\citeauthoryear{M\'{e}sz\'{a}ros et al.}{1988}]{Meszaros88} M\'{e}sz\'{a}ros P., Novick R., Chanan G.A., Weisskopf M.C., Szentgy\"{o}rgyi A., 1988, ApJ, 324, 1056
\bibitem[\protect\citeauthoryear{Morales Rueda et al.}{2002}]{Morales02} Morales Rueda L., Still M.D., Roche P., Wood J.H., Lockley J.J., 2002, MNRAS, 329, 597 
\bibitem[\protect\citeauthoryear{Nauenberg}{1972}]{Nauenberg72} Nauenberg M., 1972, ApJ, 175, 417
\bibitem[\protect\citeauthoryear{Pozdnyakov, Sobol \& Sunyaev}{Pozdnyakov et al.}{1983}]{Poz83} Pozdnyakov L.A., Sobol I.M., Sunyaev R.A., 1983, ASPRv, 2, 189
\bibitem[\protect\citeauthoryear{Press et al.}{1992}]{Press92} Press W.H., Teukolsky S.A., Vetterling W.T., Flannery B.P., 1992, Numerical Recipes in Fortran: The Art of Scientific Computing 2nd Ed., Cambridge: Cambridge University Press, p. 281
\bibitem[\protect\citeauthoryear{Rees}{1975}]{Rees75} Rees M.J., 1975, MNRAS, 171, 457
\bibitem[\protect\citeauthoryear{Saxton}{2005}]{Saxton05} Saxton C.J., Wu K., Cropper M., Ramsay G., 2005, MNRAS, 360, 1091
\bibitem[\protect\citeauthoryear{\v{S}imon}{2002}]{Simon02} \v{S}imon V., 2002, A\&A, 382, 910
\bibitem[\protect\citeauthoryear{Sobol}{1979}]{Sobol79} Sobol I.M., 1979, USSR Comput. Math. Math. Phys., 18/4, 217
\bibitem[\protect\citeauthoryear{Vrielmann, Ness \& Schmitt}{Vrielmann et al.}{2005}]{Vrielmann05} Vrielmann S., Ness J.-U., Schmitt, J.H.M.M., 2005, A\&A, 439, 287 
\bibitem[\protect\citeauthoryear{Warner}{1995}]{Warner95} Warner B., 1995, Cataclysmic Variable Stars, Cambridge: Cambridge University Press
\bibitem[\protect\citeauthoryear{Wu}{1994}]{Wu94a} Wu  K., 1994, PASA, 11, 61 
\bibitem[\protect\citeauthoryear{Wu}{1999}]{Wu99} Wu K., 1999, in Hellier C., Mukai K., eds, ASP Conf. Ser. 157, Annapolis workshop on Magnetic Cataclysmic Variables (San Francisco: ASP), p339
\bibitem[\protect\citeauthoryear{Wu}{2000}]{Wu00} Wu K., 2000, Space Science Reviews, 93, 611
\bibitem[\protect\citeauthoryear{Wu, Chanmugam \& Shaviv}{Wu et al.}{1994}]{Wu94} Wu K., Chanmugam G., Shaviv G., 1994, ApJ, 426, 664
\bibitem[\protect\citeauthoryear{Wu et al.}{2003}]{Wu03} Wu K., Cropper M., Ramsay G., Saxton C., Bridge C., 2003, Chin. J. Astron. Astrophys., 3 (Suppl.), 235


\end{thebibliography}
\end{document}